# The ELT-MOS (MOSAIC): towards the construction phase

Simon Morris*[a], François Hammer[b], Pascal Jagourel[b], Christopher J. Evans[c],; Mathieu Puech[b], Gavin B. Dalton[c], Myriam Rodrigues[b], Ruben Sanchez-Janssen[c], Ewan Fitzsimons[c], Beatriz Barbuy[d], Jean-Gabriel Cuby[e], Lex Kaper[f], Martin Roth[g], Gérard Rousset[b], Richard Myers[a], Olivier Le Fèvre[e], Alexis Finogenov[h], Jari Kotilainen[i], Bruno Castilho[j], Goran Ostlin[k], Sofia Feltzing[l], Andreas Korn[m], Jesus Gallego[n], África Castillo Morales[n], Jorge Iglesias-Páramo[o], Laura Pentericci[p], Bodo Ziegler[q], Jose Afonso[r], Marc Dubbledam[a], Madeline Close[a], Phil Parr-Burman[c], Timothy J. Morris[a], Fanny Chemla[b], Fatima De Frondat[b], Andreas Kelz[g], Isabelle Guinouard[b], Ian J. Lewis[s], Kevin Middleton[c], Ramon Navarro[t], Marie LarrieuJohan Pragt[u], Annemieke Janssen[t], Kjetil Dohlen, Kacem El Hadi[e], Éric Gendron[b], Yanbin Yang[b], Martyn Wells[c], Jean-Marc Conan[v], Thierry Fusco[v], Daniel Schaerer[w], Edwin Bergin[x], Sylvestre Taburet[b], Mickaël Frotin[b], Nadia Berkourn[b].
[a]Physics Department, Durham University(UK); [b]Observatoire de Paris (France); [c]Science and Technology Facilities Council (United Kingdom); [d]Univ. de São Paulo (Brazil); [e]Lab. d'Astrophysique de Marseille (France); [f]Univ. van Amsterdam (Netherlands); [g]Leibniz-Institut für Astrophysik Potsdam (Germany); [h]Univ. of Helsinki (Finland); [i]FINCA - Finnish Ctr. for Astronomy with ESO (Finland); [j]Lab. Nacional de Astrofísica (Brazil); [k]Stockholm Univ. (Sweden); [l]Lund Univ. (Sweden); [m]Uppsala Univ. (Sweden); [n]Univ. Complutense de Madrid (Spain); [o]Instituto de Astrofísica de Andalucía - CSIC (Spain); [p]INAF - Osservatorio Astronomico di Roma (Italy); [q]Univ. Wien (Austria); [r]Observatório Astronómico de Lisboa (Portugal); [s]Univ. of Oxford (United Kingdom); [t]Netherlands Research School for Astronomy (Netherlands); [u]Institut de Recherche en Astrophysique et Planétologie (France); [v]ONERA (France); [w]Université de Genève (Switzerland); [x]University of Michigan (USA)

## ABSTRACT

When combined with the huge collecting area of the ELT, MOSAIC will be the most effective and flexible Multi-Object Spectrograph (MOS) facility in the world, having both a high multiplex and a multi-Integral Field Unit (Multi-IFU) capability. It will be the fastest way to spectroscopically follow-up the faintest sources, probing the reionisation epoch, as well as evaluating the evolution of the dwarf mass function over most of the age of the Universe. MOSAIC will be world-leading in generating an inventory of both the dark matter (from realistic rotation curves with MOAO fed NIR IFUs) and the cool to warm-hot gas phases in z=3.5 galactic haloes (with visible wavelenth IFUs). Galactic archaeology and the first massive black holes are additional targets for which MOSAIC will also be revolutionary. MOAO and accurate sky subtraction with fibres have now been demonstrated on sky, removing all low Technical Readiness Level (TRL) items from the instrument. A prompt implementation of MOSAIC is feasible, and indeed could increase the robustness and reduce risk on the ELT, since it does not require diffraction limited adaptive optics performance. Science programmes and survey strategies are currently being investigated by the Consortium, which is also hoping to welcome a few new partners in the next two years.

*simon.morris@durham.ac.uk

## 1. INTRODUCTION

The ELT will be the world largest visible/NIR telescope in the 2020s, with a collecting area similar to the total of the 16 currently existing 8-10 meter class telescopes. Multi-object spectroscopy will add efficiency to its unprecedented sensitivity, to provide unique breakthroughs which cannot be obtained by any other means. Delivering such an instrument is why the MOSAIC Consortium has gathered a unique collection of world-wide expertise on multiple-object spectrographs, integral-field units, and adaptive optics assisted instruments. Team members conceived, designed, built and delivered on sky almost all the spectrographs currently operating at the VLT. We review the status of the MOSAIC instrument after completion of the instrument Phase A below. The last comprehensive summary of the MOSAIC instrument was given by Hammer et al. in 2016[1].

The Phase A Consortium includes countries in the core of the project (Partners: France, United Kingdom, The Netherlands, Brazil, Germany) as well as other participating countries (Associate Partners: Austria, Finland, Italy, Portugal, Spain, Sweden). This includes those interested in financially supporting the project in exchange of Guaranteed Observing Time, or in participation in possible Public Surveys.

## 2. SCIENCE CASES

During the Phase A study, the MOSAIC Consortium gathered with the science community at large in three key international meetings (Cefalu, September 2015; Paris, March 2016; Toledo, October 2017), and prioritized the numerous science cases developed over the years for MOSAIC, and which are expected to open many new astronomical avenues in the coming decade[2,3,4,5,6]. These discussions identified four key science cases which have been used to help with instrument tradeoffs:

| Survey & related Science Case(s) (not in priority order) | Primary/essential (secondary/beneficial) modes |
|---|---|
| 1. Evolution of dwarf galaxies SC2: mass assembly; SC1: contribution to reionisation | HDM (HMM-NIR) |
| 2. Inventory of matter SC2: IGM tomography; SC2: missing baryons; SC3: dark matter profiles in high-z galaxies | HDM + HMM-Vis (VIFU) |
| 3. First-light galaxies SC1: the sources of reionisation | HMM-NIR (HDM) |
| 4. Extragalactic stellar populations SC5: evolved populations beyond the Local Group | HDM (HMM-NIR) |

### 2.1 Evolution of Dwarf Galaxies

Advancing the field of galaxy formation requires a comprehensive census of the mass assembly and star-formation histories of distant galaxies. In particular, it is important to obtain spatially-resolved chemo-dynamical information for $1<z<4$ galaxies across a wide range of masses (dwarfs to giants) and environments. Such observations will allow us to infer the amount of rotational support of the galaxy, to constrain the properties and dynamics of the stellar populations and the ISM, and to derive the distribution of metals. Mapping out the start formation history of these objects will greatly improve our understanding of the contributors to the UV background as a function of epoch. Only by building large statistical samples of high-z systems, that are characterised to this level of detail, will we be able to constrain the physics of galaxy formation.

### 2.2 Inventory of matter (including mass assembly)

The gas in the IGM is revealed by the numerous hydrogen absorption lines that are seen in the spectra of quasars bluewards of the Ly-α emission from the quasar. It has been shown that the high-z IGM contains most of the baryons in the Universe and is therefore the baryonic reservoir for galaxy formation. In turn, galaxies emit ionising photons and expel metals and energy through powerful winds that determine the physical state of the gas in the IGM. This interplay of galaxies and gas is central to the field of galaxy formation. It is complex and happens on scales of the order of 1 Mpc (~2 arcmin on the sky at z~2.5, using standard cosmological parameters). The main goal of this case is to reconstruct the 3D density field of the IGM at z~2.5 on these and larger scales to study its topology, its chemical properties, and to correlate the position of the galaxies with the gas density peaks.

### 2.3 First Light Galaxies

Exactly when and how reionisation occurred is still poorly constrained. Identification of the main (and elusive) ionising sources requires detailed observations of the Ly-α properties of faint objects. This will enable a precise characterisation of the ionisation state of the intergalactic medium (IGM) in the first Gyr ($5 < z < 13$), allowing us to reconstruct the timeline of reionisation. MOSAIC will determine the properties of the first galaxies, including their stellar populations, their interstellar media (ISM) and the presence of outflows.

**2.4 Extragalactic Stellar Populations**

Resolved stellar populations enable us to explore the star-formation and chemical-enrichment histories of galaxies, providing direct constraints on galaxy formation and evolution models. Studies in the Local Group have shown that precise chemical abundances and stellar kinematics can break the age-metallicity degeneracy, while also helping to disentangle the populations associated with different structures. However, to fully assess the diversity of galaxy populations we need to move to a broader range of galaxies in the Local Volume, from the edge of the Local Group out to Mpc distances. As we move to larger distances, observations of the evolved stellar populations will move from 'resolved' to 'semi-resolved' in terms of spatial resolution, building a bridge toward integrated-light studies. These programmes will include studies of extragalactic star clusters, starbursts and nuclear activity in merging galaxies, assessing e.g. invariance of the initial mass function and characterising different structural components of external galaxies.

**2.5 Serendipity and Parameter Space**

All of the astronomers involved in the MOSAIC consortium are very aware of the history of astronomical instrumentation, whereby the breakthroughs produced by an instrument are rarely the science cases developed up to 10 years before the delivery of that instrument. Guided by this experience, preserving instrument discovery space, and flexibility has always been an additional consideration during instrument design trade-offs.

## 3. FROM SCIENCE TO TOP LEVEL REQUIREMENTS TO DESIGN

During the Phase A study, we performed early trade-offs which were applied to the so-called 'full' MOSAIC option (itself already a compromise between cost and capability), which must be sufficiently versatile to address the community identified scientific priorities. The full MOSAIC includes high multiplex modes (HMM) in the visible and NIR, high definition IFUs in the NIR (HDM), and telescope PSF limited IFUs in the visible (IGM renamed VIFU) modes (see section 4 below). The primary design goals are then:
 - To keep the system manageable. (For example, given the large spectral coverage, the different modes, and the foreseen multiplex, we require a carefully designed system to link the focal plane and the spectrograph);
 - To ensure the instrument is competitive in terms of throughput and survey efficiency.

The latter has led us to limit the wavelength range to 450-1800 nm, because of the expected performance of the ELT mirror coatings in the far blue, and the expectedly strong background from the 5-mirror telescope in the K band, respectively. In the full MOSAIC option with 5 NIR and 5 VIS spectrographs, the multiplex is 200 in the visible, 100 in the NIR (with paired fibres for precision sky-subtraction purposes) and, finally, a multiplex of 10 for the HDM IFUs.

## 4. PHASE A DESIGN

To satisfy the assembled science cases and requirements, the Phase A design includes four observational modes:

**High-definition mode (HDM)**: Simultaneous observations of 8 IFUs (goal: 10 IFUs) deployed within a ~40 arcmin$^2$ patrol field and each with enhanced image quality from multi-object adaptive optics. Each IFU will cover a 1.9 arcsec hexagon with 80 mas spaxels, with the spectrographs delivering R~5000 over 0.8-1.8μm (between 250 and 430nm in one observation). A high spectral-resolution set-up will provide R~20,000 over a passband of ~100nm at ~1.6μm. The light entering each spaxel will be injected into a fibre that will be imaged onto 3.06 pixels on the detector.

**Visible Integral Field Unit mode (VIFU)**: Simultaneous observations of 8 IFUs (goal: 10 IFUs) deployed within a ~40 arcmin$^2$ patrol field with telescope delivered 'seeing-limited' performance (or GLAO-corrected, using M4 of the telescope). Each IFU will cover a 2.3 arcsec hexagon with 138 mas spaxels, with the spectrographs delivering R~5000 over 0.45-0.92μm (between 150 and 230nm in one observation). A possible high spectral-resolution set-up with

R~15,000 could allow coverage of a passband of ~60nm around 0.65μm and/or 0.86μm. The light entering each spaxel will be injected into a fibre that will be imaged onto 4.21 pixels on the detector.

**High multiplex mode in the near-IR (HMM-NIR)**: Simultaneous integrated-light observations of 80 objects (goal: 100) with dedicated sky fibres operated in cross-beam switching sequences. Each object will be observed with a bundle of 19 x 100 mas fibres, giving an on-sky aperture of 500 mas in diameter. Each fibre will be imaged onto 3.06 pixels on the detector. The spectrograph setups are identical to the HDM mode.

**High multiplex mode in the visible (HMM-Vis)**: Simultaneous integrated-light observations of 160 objects (goal: 200). Note that there will be no dedicated sky fibres in this mode, as the sky background is much fainter than in the near-IR. Each object will be observed with a bundle of 19 x 168 mas fibres, giving an on-sky aperture of 840 mas in diameter. Each fibre will be imaged onto 4.21 pixels on the detector. The spectrograph setups are identical to the VIFU mode.

Figure 1 summarises the modular architecture of the instrument, and the 4 observational modes delivered by MOSAIC.

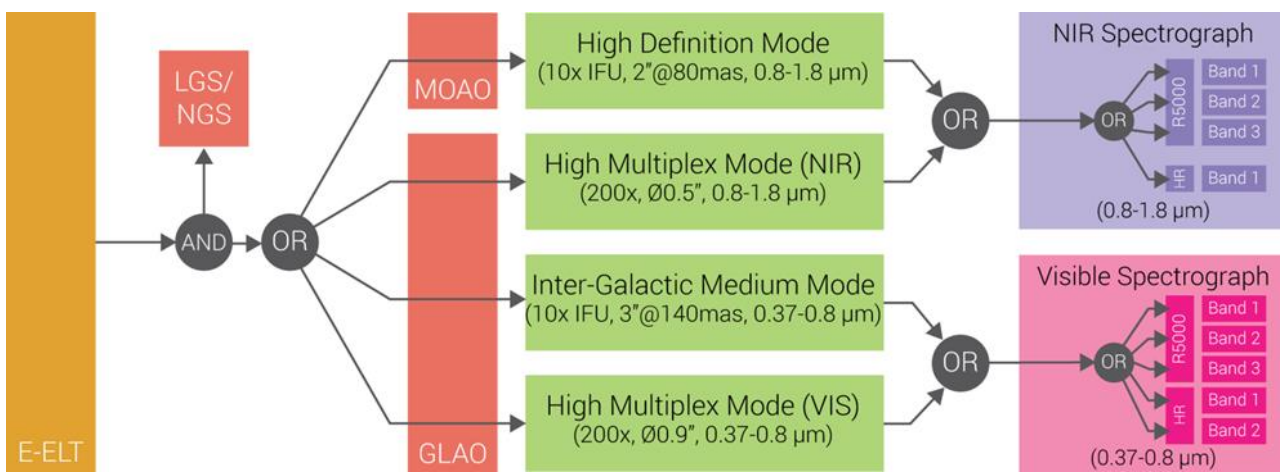

Figure 1. The MOSAIC instrument concept. Note that various descope options are still being considered (see section 5).

MOSAIC has been designed to be highly modular, allowing for potential descopes and phased deployment. Figure 2 illustrates this.

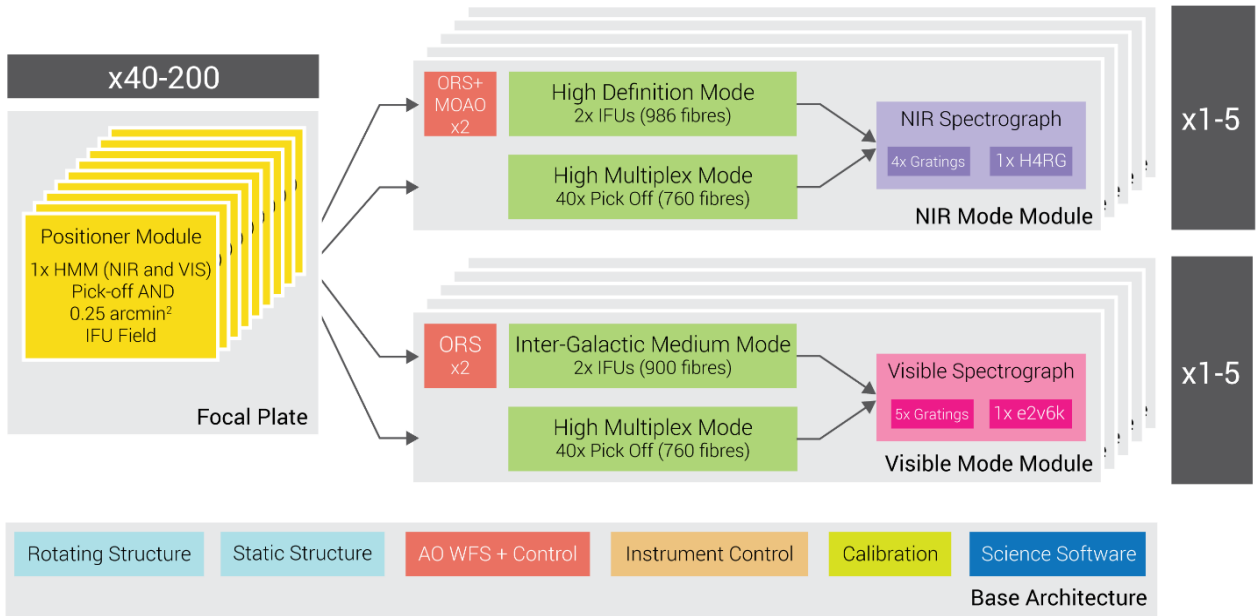

Figure 2. The MOSAIC modular approach, allowing for both a possible phased delivery and/or descope options.

As the focal plane of the instrument will be rotating around the telescope azimuth axis, the gravitational loads on the instrument core depend on the orientation of the instrument. The rotating structure will therefore be subjected to significant non-axial loads, inducing substantial out-of-plane bending moments. In order to achieve an acceptable resistance to the gravity-induced deformations, the design shown in the Figure 3 has been adopted.

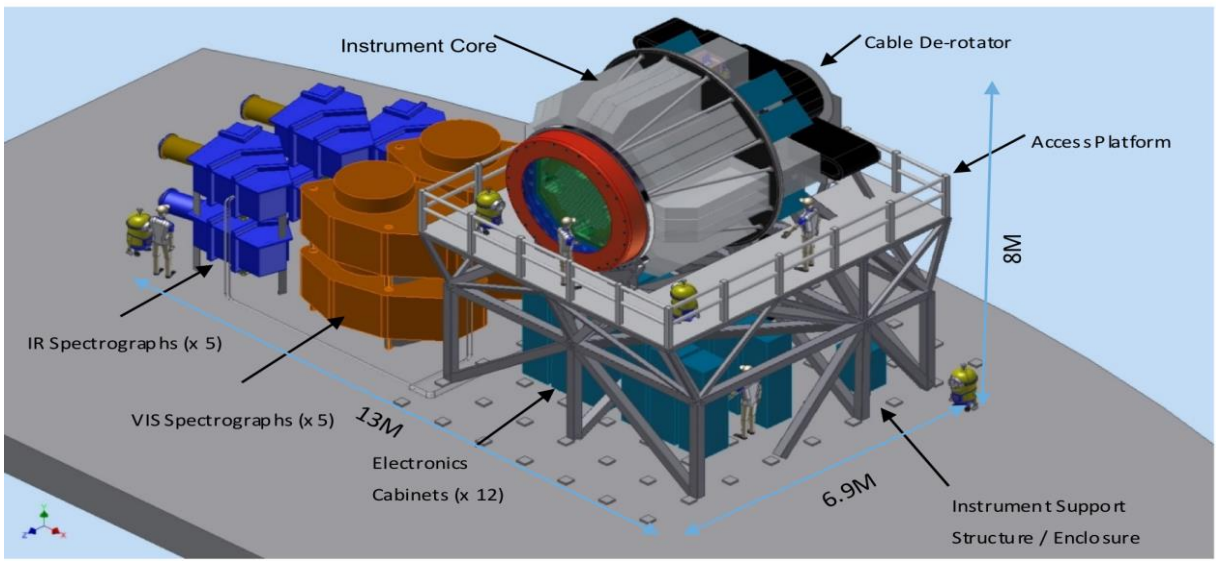

Figure 3. The MOSAIC mechanical design. Note that various descope options are still being considered (see section 5).

## 5. THE NEXT STEPS

The funding context has evolved since the start of the MOSAIC Phase A in early 2016. The expected 18 M€ for hardware defined in the original SoW cannot currently be guaranteed on a reasonable timescale, since the full ELT Phase 1 and Phase 2 costs have not yet been covered by savings and additional cash from new ESO member states. This prompted ESO to contact the PIs (MOS & HIRES), resulting in two face-to-face meetings in 2017 to discuss and agree on common actions for preparing an "agreement to construction", under a scheme whereby the Consortia gather cash for the hardware in exchange for larger GTO rewards. This approach was recently approved by the ESO Council.

During the Phase A, various descope options were also explored. These were in turn discussed with the MOSAIC steering committee, who represent the current partners who may expect to make a contribution to the hardware costs. A descope option delivering an HMM mode with a multiplex of 80-100, and an HDM mode with a multiplex of 8-10 in both the visible and NIR was considered acceptable, if the cost/mass/volume limits require this.

A bottom up estimate of the times for each phase of the instrument design and construction led to an estimate that, given a Phase B start in early 2019, then instrument acceptance on the telescope could be in 2027.

Having successfully completed the instrument Phase A, the MOSAIC consortium is now working closely with ESO staff to deliver the instrument as soon as possible after the telescope first light, both to enable unique science, and also to improve the ELT robustness and efficiency with a world class multi-object spectrograph capability.

Considerably more detail about many aspects of the MOSAIC instrument were presented separately at this SPIE Conference[7,8,9,10,11,12,13,14,15].